\documentclass[floatfix,pra,superscriptaddress,twocolumn]{revtex4}
\usepackage{optidef}

\usepackage{graphicx,bm,natbib,upgreek,amsmath,mathrsfs,accents}
\usepackage{amsbsy}
\usepackage{amsfonts}

\usepackage{amsthm}

\usepackage[dvipsnames]{xcolor}

\usepackage{graphicx}
\usepackage{subfigure}

\usepackage{hyperref}

\usepackage[nice]{nicefrac}

\usepackage[normalem]{ulem}

\usepackage{color}

\begin{document}

\title{Single-photon sub-Rayleigh precision measurements \\ of a pair of incoherent sources of unequal intensity}

\author{Luigi Santamaria}
\author{Fabrizio Sgobba}
\affiliation{Agenzia Spaziale Italiana, Matera Space Center, Contrada Terlecchia snc.~75100 Matera, Italy}

\author{Cosmo Lupo}
\affiliation{Dipartimento Interateneo di Fisica, Politecnico \& Universit\`a di Bari, 70126, Bari, Italy}
\affiliation{INFN, Sezione di Bari, 70126 Bari, Italy}


\begin{abstract}
Interferometric methods have been recently investigated to achieve sub-Rayleigh imaging and precision measurements of faint incoherent sources up to the ultimate quantum limit.
Here we consider single-photon imaging of two point-like emitters of unequal intensity. This is motivated by the fact that pairs of natural emitters will typically have unequal brightness, as for example binary star systems and exoplanets.
We address the problems of estimating the transverse separation $d$ and the relative intensity $\epsilon$.
Our theoretical analysis shows that the associated statistical errors are qualitatively different from the case of equal intensity.
We employ multi-plane light conversion technology to implement Hermite-Gaussian (HG) spatial-mode demultiplexing (SPADE), and demonstrate sub-Rayleigh measurement of two emitters with Gaussian point-spread function. The experimental errors are comparable with the theoretical bounds. 
The latter are benchmarked against direct imaging, yielding a $\epsilon^{-1/2}$ improvement in the signal-to-noise ratio, which may be significant when the primary source is much brighter than the secondary one, as for example for imaging of exoplanets.
However, achieving this improved scaling requires low noise in the implementation of SPADE, which is typically affected by crosstalk between HG modes.
\end{abstract}

\maketitle

\section{Introduction}

What is the ultimate limit of optical resolution?~According to the well-known Rayleigh criterion~\cite{Rayleigh1879}, a diffraction-limited imaging system can resolve two point-like emitters as long as their transverse separation is larger than or comparable to the Rayleigh length $\mathrm{x_R} = \lambda D/R$, where $\lambda$ is the wavelength, $R$ is the radius of the pupil, and $D$ the distance to the emitters. The criterion follows from the fact that, due to diffraction through the pupil, the image of a point-source is a spot of size $\mathrm{x_R}$, determined by the point-spread function (PSF) characterizing the optical system~\cite{goodman2008introduction}.
Ways to circumvent this fundamental limit have been developed over time, one of the most important being super-resolved fluorescence microscopy~\cite{Hell1994},
which makes use of non-linear optics to ensure that nearby emitters do not emit at the same time. 
Within quantum imaging~\cite{kolobov2007quantum}, methods have been proposed that exploit the fact that a quantum state of light with exactly $N$ photons has an effective wavelength that is $N$ times shorter~\cite{JPD}, which in turn implies a PSF $N$ times sharper~\cite{Giovannetti}.

Note that the above strategies rely on source engineering, which is not always an option, as for example in astronomic observations.
Furthermore, they are based on direct imaging (DI), i.e., pixel-by-pixel measurement of the intensity of the field focused on the image plane.
However, DI ignores the information carried by the phase of the field, which instead can be uncovered by interferometric measurements. 
The latter may allow us to demultiplex the field in the image plane and measure it in a non-local basis of optical modes obtained as superposition of the field in different pixels. With this approach one may hope to achieve the ultimate quantum limit of detection~\cite{Helstrom_book}.

In 2015, Tsang, Nair, and Lu~\cite{Tsang2016} framed the problem of finding the ultimate far-field optical resolution as a problem of quantum estimation~\cite{Paris,Sidhu}. 
They considered the estimation of the median point and transverse separation of two point-like emitters of equal intensity.
For finite number of photons detected, the statistical error is bounded from below by the Cramér-Rao bound, which in turn is expressed in terms of the Fisher information. 
One can show that for DI, the Fisher information approaches zero when the separation becomes smaller than the Rayleigh length, which implies large statistical errors: a manifestation of the Rayleigh resolution criterion.
In contrast, interferometric measurements show a different behavior for the Fisher information.
Spatial-mode demultiplexing (SPADE) of the field in the image plane can be optimized, depending on the particular form of the PSF, to yield a constant Fisher information, independently of the value of the transverse separation.
As a matter of fact, in the regime of single-photon imaging, SPADE (with other interferometric techniques ~\cite{Nair:16}) has been shown to be the globally optimal measurement, for this and for similar estimation problems~\cite{Tsang2016,Dutton,PRL2020}, among the broader set of all measurements compatible with the principles of quantum mechanics.
The optimality of SPADE is not confined to estimation theory though, and applications have been explored to the related problem of hypothesis testing~\cite{lu2018quantum,PRL2021,Sorelli}, and beyond the case of two point-sources~\cite{GraceGuha}.

The work of Tsang et al.~sparkled a renewed interest in quantum imaging, which in turn led to a number of theoretical and experimental results, see Ref.~\cite{Trev} and the website of Tsang's group~\footnote{\href{https://blog.nus.edu.sg/mankei/superresolution/}{https://blog.nus.edu.sg/mankei/superresolution/}} for a list of contributions. 
In practice, SPADE can be implemented experimentally in a number of ways, e.g.~with spatial light modulators~\cite{Par2016,Paur:18,Zhou2019,Zhou:23}, by image inversion~\cite{Tham2017}, with a photonic lantern~\cite{Salit:20}, and through multi-plane light conversion~\cite{Boucher2020,Santamaria22,Treps23}.
While SPADE has been proven optimal (in an information-theoretic sense) in the regime of highly attenuated signals with at most one photon per detection~\cite{Tsang2016}, the same approach may yield sub-Rayleigh resolution in the regime of bright sources, suggesting quantum-inspired methods for imaging and precision measurement~\cite{Lvovsky,Boucher2020,Santamaria22,Lvovsky2023}.
The advantage of SPADE over DI may persist, with some subtle modifications and if the transverse separation is not too small, even in the presence of noise and crosstalk, as a long as they are not too strong~\cite{Gessner2020,Banaszek,PhysRevA.101.022323,ct2022}.
Recently, a modern version of the Rayleigh criterion has been formulated based on this approach~\cite{Modern}, which takes into account the information contained in the phase that can be extracted through SPADE~\cite{PhysRevA.99.012305,PhysRevA.104.052411,TsangIII}.


In this paper, we report an experimental demonstration of sub-Rayleigh precision measurement, obtained through SPADE in the single-photon regime.
Our experiment simulates a pair of ultra-weak incoherent point-sources that are observed through a diffraction-limited optical system characterized by its Rayleigh length. 
Unlike previous experimental work~\cite{Par2016,Paur:18,Zhou2019,Zhou:23,Tham2017,Treps23}, here we consider sources of different intensities. This is motivated by the fact that pairs of natural emitters will typically have unequal brightness.
In particular, this may have potential application for the observation of exoplanets~\cite{PhysRevA.96.062107,PRL2021,astro22}.
We implement SPADE using a telecom-wavelength demultiplexer from Cailabs~\footnote{\href{https://www.cailabs.com/en/technology/}{https://www.cailabs.com/en/technology/}}, which sorts the optical field in the image plan in the basis of Hermite-Gaussian (HG) modes.
In an independent work by Rouvière et al.~\cite{Treps23}, this demultiplexer has been used to estimate the spatial separation between two equally bright sources~\footnote{The paper by Rouvière et al.~\cite{Treps23} appeared online after we completed our measurements, while we were writing the present manuscript.}.
As we discuss here, the error analysis is qualitatively different when the source have unequal intensity, see also Refs.~\cite{PhysRevA.96.062107,ct2022}, and crucially depends on how the optical system is aligned.
Our scheme exploits only two photon detectors, which are used to count photons in the lower-order modes $\text{HG}_{01}$ and $\text{HG}_{10}$. This is sufficient to estimate the distance $d$ between the sources or, in alternative, their relative intensity $\epsilon$. 
The experiment is supported by theoretical modeling.
We compute theoretical bounds on the statistical errors and compare with DI.
In particular, we show that in principle noiseless SPADE can improve the signal-to-noise (SNR) ratio by a factor $\epsilon^{-1/2}$, which may be significant when the primary source is much brighter than the secondary one, as for example for imaging of exoplanets.


\section{The model}

Our experimental setup consists of two sources separated by a distance $d$, with relative intensities $\epsilon$ and $1-\epsilon$.
We consider the limit of faint signals, where most of the times no photon is detected and the field is in the vacuum state $|0\rangle$. The probability of detecting a photon is $\eta \ll 1$ and the probability of multiple photon events is negligibly small.
In this regime, the state of a single spatio-temporal mode of the field is represented by a density matrix of the following form,
\begin{align}\label{rho0}
\rho = (1-\eta) |0\rangle \langle 0|
+ \eta (1-\epsilon) |\psi_0\rangle \langle \psi_0|
+ \eta \epsilon |\psi_d\rangle \langle \psi_d| \, ,
\end{align}
where $\psi_0$ and $\psi_d$ are the states of the photon emitted by either source, and $d$ is their transverse separation. 
In the single-photon limit, $(1-\epsilon)$ and $\epsilon$ are the probabilities that the photon populates the state $\psi_0$ or $\psi_d$, respectively. 
Therefore, for $\epsilon < 1/2$, the state $\psi_0$ is the image of the brighter source.
If the brighter source has intensity $I_A$ and the weaker source has intensity $I_B$, then $\epsilon \simeq I_B/I_A$ when $I_B \ll I_A$. 

In our experimental setup, we align the optical system to the brighter source. This corresponds to centering the state $\psi_0$ in the origin of the reference system in the image plane, whereas the $\psi_d$ is centered at position $(d_x,d_y)$, with $d= \sqrt{d_x^2 + d_y^2}$.
For a Gaussian PSF of width $w_0$ we have
\begin{align}
    |\psi_0\rangle & = \mathcal{N} \int d\mu \,
    e^{-\frac{x^2+y^2}{4 w_0^2}} a^\dag(x,y) |0\rangle \, , \\
    |\psi_d\rangle & = \mathcal{N} \int d\mu \,
    e^{-\frac{(x-d_x)^2+(y-d_y)^2}{4 w_0^2}} a^\dag(x,y) |0\rangle \, ,
\end{align}
with normalization factor $\mathcal{N} = (2\pi w_0^2)^{-1/2}$ and $d\mu = dx \, dy$.
Here $a^\dag(x,y)$ is the canonical bosonic operator associated to a photon detection at position $(x,y)$ in the image plane.
Such a state of the field is measured in the HG modes, i.e., the field is first demultiplexed in the HG basis in the image plane, then measured by photon detection.
We focus on the lower-order terms, which are represented by projections on the HG normalized modes,
\begin{align}
    |\text{HG}_{00}\rangle & = \mathcal{N} \int d\mu \, e^{-\frac{(x-\delta_x)^2+(y-\delta_y)^2}{4 w_0^2}} a^\dag(x,y) |0\rangle \, , \\
    |\text{HG}_{01}\rangle & = \mathcal{N} \int d\mu \, \frac{x-\delta_x}{w_0} \, e^{-\frac{(x-\delta_x)^2+(y-\delta_y)^2}{4 w_0^2}}  a^\dag(x,y) |0\rangle \, , \\
    |\text{HG}_{10}\rangle & = \mathcal{N} \int d\mu \, \frac{y-\delta_y}{w_0} \, e^{-\frac{(x-\delta_x)^2+(y-\delta_y)^2}{4 w_0^2}}  a^\dag(x,y) |0\rangle \, ,
\end{align}
where $\delta_x$ and $\delta_y$ account for residual misalignment. 
Ideally, for $\delta_x = \delta_y = 0$, the interferometer is perfectly aligned towards the brighter source.


\subsection{Alignment}\label{ssec:align}

In our experimental setup, we align the optical system to the brighter source. 
This is in contrast to what is done in other works, where the alignment is on the centroid or median point~\cite{Tsang2016,Nair:16,Paur:18,Par2016,GraceGuha,Treps23}.
To align the optical system, we maximize the signal in mode $\text{HG}_{00}$. 
In fact, the probability of detecting a photon in the lower HG mode is
\begin{align}
    p_{0} & = \langle \text{HG}_{00} | \rho | \text{HG}_{00} \rangle \\
    & = \eta (1-\epsilon) | \langle \text{HG}_{00}|\psi_0\rangle |^2
    + \eta \epsilon | \langle \text{HG}_{00}|\psi_d\rangle |^2 \\
    & = \eta (1-\epsilon) e^{ - \frac{\delta_x^2 +\delta_y^2 }{4 w_0^2}}
    + \eta \epsilon e^{ - \frac{(d_x-\delta_x)^2 + (d_y-\delta_y)^2 }{4 w_0^2}} \, .
    \label{p00}
\end{align}

In the sub-Rayleigh regime (when $d$ is much smaller than $w_0$) we approximate
\begin{align}
    p_{0} & \simeq \eta (1-\epsilon) \left( 1 - \frac{\delta_x^2+\delta_y^2}{4 w_0^2} \right)
    \nonumber \\
    & \phantom{=}~
    + \eta \epsilon \left( 1 - \frac{(d_x-\delta_x)^2+(d_y-\delta_y)^2}{4 w_0^2} \right) \\
    & \simeq \eta - \eta (1-\epsilon) \frac{\delta_x^2+\delta_y^2}{4 w_0^2} 
    - \eta\epsilon \frac{(d_x-\delta_x)^2+(d_y-\delta_y)^2}{4 w_0^2} \, .
\end{align}
Experimentally, the alignment is performed when the two sources are super-imposed, i.e., for $d_x=d_y=0$. In which case we obtain
\begin{align}
    p_{0} & \simeq \eta - \eta \frac{\delta_x^2+\delta_y^2}{4 w_0^2} 
    \, .
\end{align}
In general, this approach may be used even when the sources are not superimposed, as long as $\epsilon$ is small enough.
In fact the maximum of $p_{0}$ is in general reached when
$\delta_x = \epsilon d_x$ 
and 
$\delta_y = \epsilon d_y$.


\subsection{Theoretical error bounds}\label{ssec:CRB}

Our main goal is either the estimation of the separation $d$ between the two sources (if the relative intensity is known), or their relative intensity $\epsilon$ (if the separation is known). In order to achieve this goal we measure the field in the lower HG modes, $\text{HG}_{00}$, $\text{HG}_{01}$, $\text{HG}_{10}$. 
In addition to $p_{0}$ in Eq. (\ref{p00}), we have
\begin{align}
    p_{01} & = \langle \text{HG}_{01} | \rho | \text{HG}_{01} \rangle \\
    & =
    \eta (1-\epsilon) \frac{\delta_x^2}{4 w_0^2}
    e^{-\frac{\delta_x^2+\delta_y^2}{4 w_0^2}}
    \nonumber\\
    & \phantom{=}~ + \eta \epsilon \frac{(d_x-\delta_x)^2}{4 w_0^2}
    e^{-\frac{(d_x-\delta_x)^2+(d_y-\delta_y)^2}{4 w_0^2}}
    \, , \\
    p_{10} & = \langle \text{HG}_{10} | \rho | \text{HG}_{01} \rangle \\
    & =
    \eta (1-\epsilon) \frac{\delta_y^2}{4 w_0^2}
    e^{-\frac{\delta_x^2+\delta_y^2}{4 w_0^2}}
    \nonumber\\
    & \phantom{=}~
    + \eta \epsilon \frac{(d_y-\delta_y)^2}{4 w_0^2}
    e^{-\frac{(d_x-\delta_x)^2+(d_y-\delta_y)^2}{4 w_0^2}} \, .
\end{align}

It is useful to consider the joint events of detection in either the $\text{HG}_{01}$ or $\text{HG}_{10}$ modes, which happens with probability
\begin{align}
    p_1 & = p_{01} + p_{10} \\
    & = \eta (1-\epsilon) \frac{\delta_x^2+\delta_y^2}{4 w_0^2}
    e^{-\frac{\delta_x^2+\delta_y^2}{4 w_0^2}}
    \nonumber\\
    & \phantom{=}~
    + \eta \epsilon \frac{(d_x-\delta_x)^2+(d_y-\delta_y)^2}{4 w_0^2}
    e^{-\frac{(d_x-\delta_x)^2+(d_y-\delta_y)^2}{4 w_0^2}} \, .
\end{align}
In the sub-Rayleigh regime this reads
\begin{align}
    p_1 & \simeq \eta (1-\epsilon) \frac{\delta_x^2+\delta_y^2}{4 w_0^2} + \eta \epsilon \frac{(d_x-\delta_x)^2+(d_y-\delta_y)^2}{4 w_0^2}
    \, .
\end{align}

Ideally, for perfect alignment we would have $\delta_x=\delta_y=0$. However, in a practical experimental setup we expect a residual misalignment. Whereas the alignment procedure aims at reducing the systematic error due to misalignment, we model the residual misalignment as a stochastic error. We assume that $\delta_x$ and $\delta_y$ are stochastic variables with zero mean and variance $\sigma^2/2$:
\begin{align}
    \langle \delta_x \rangle = \langle \delta_y \rangle & = 0 \, , \\
    \langle \delta_x^2 \rangle = \langle \delta_y^2 \rangle & = \sigma^2/2 \, .
\end{align}
This assumption in turn yields
\begin{align}
    p_{0} & = 
\eta - \eta \frac{\sigma^2}{4 w_0^2} 
    - \eta\epsilon \frac{ d^2 }{4 w_0^2} \, , \label{p0} \\
    p_1 & =
\eta \frac{\sigma^2}{4 w_0^2} + \eta \epsilon \frac{d^2}{4 w_0^2} \, . \label{p1}
\end{align}


Our strategy is to first estimate the re-scaled probability $p_1/\eta$. Then, from $p_1/\eta$, we obtain either $d$ or $\epsilon$.
Experimentally, we may measure the relative frequency $f_1 = n_1/(n_0 + n_1)$, where $n_0$ and $n_1$ are the number of detection in modes $\text{HG}_{00}$ and $\text{HG}_{01}+\text{HG}_{10}$, respectively.
Note that $f_1$ is an unbiased estimator for $p_1/\eta$, where $n_0 + n_1 = \eta n$.
The variance is obtained from the binomial distribution:
\begin{align}
\Delta f_1 
= \frac{1}{\eta} \sqrt{ \frac{p_1 p_0}{\eta n} } 
%
\simeq  
\sqrt{ \frac{ 
\frac{\sigma^2}{4 w_0^2} + \epsilon \frac{d^2}{4 w_0^2} 
} {\eta n} } \, .
\end{align}

In turn, the variance for the estimation of $d$ and $\epsilon$ is obtained by error propagation:
\begin{align}
\Delta d 
= \left( \frac{\partial p_1/\eta }{\partial d } \right)^{-1 } \Delta f_1  
%
\simeq \frac{2 w_0^2}{\epsilon d}
\sqrt{ \frac{ 
\left( \frac{\sigma^2}{4 w_0^2} + \epsilon \frac{d^2}{4 w_0^2} \right) 
} {\eta n} } \, .
\end{align}
Similarly
\begin{align}
\Delta \epsilon 
= \left( \frac{\partial p_1/\eta }{\partial \epsilon} \right)^{-1 } \Delta f_1  
%
\simeq \frac{4 w_0^2}{d^2} 
\sqrt{ \frac{ 
\frac{\sigma^2}{4 w_0^2} + \epsilon \frac{d^2}{4 w_0^2} 
} {\eta n} } \, .
\end{align}


\subsection{Crosstalk}\label{ssec:ct}

Ideally, SPADE would exactly project the field into the HG modes.
However, in practice demultiplexing is affected by crosstalk.
As our focus is on the lower order modes $\text{HG}_{00}$, $\text{HG}_{01}$, $\text{HG}_{10}$, the crosstalk is characterized by the probability $\chi$ that a photon in the 
$\text{HG}_{00}$ mode (which is brighter) is detected in either the $\text{HG}_{01}$ or $\text{HG}_{10}$ modes.
Therefore, the probabilities in Eqs.~(\ref{p0})-(\ref{p1}) need to be replaced by
\begin{align}
    p_{0} & \to
               (1-\chi) p_0
                 + \chi p_1 
          \simeq 
               (1-\chi) \eta 
               - \eta \frac{\sigma^2}{4 w_0^2} - \eta \epsilon \frac{d^2}{4 w_0^2} \, , \\
    p_1 & \to
               (1-\chi) p_1
               + \chi p_0
        \simeq    
               \chi \eta + \eta \frac{\sigma^2}{4 w_0^2} + \eta \epsilon \frac{d^2}{4 w_0^2} \, .
\end{align}
In turn, the effect of crosstalk can then be included in the above model by replacing $\sigma^2$ with $\sigma^2 + 4 w_0^2 \chi$,  and the statistical errors read
\begin{align}
\Delta d 
& = w_0 \sqrt{ \frac{ \frac{\sigma^2 + 4 w_0^2 \chi}{\epsilon d^2} + 1 }{ \eta n \epsilon}} \, , 
\label{errd} \\
\Delta \epsilon
& = \frac{2 w_0}{d} 
\sqrt{\frac{ \frac{\sigma^2 + 4 w_0^2 \chi}{\epsilon d^2} + 1 }{\eta n /\epsilon }}
\, .
\label{erre}
\end{align}

In conclusion, this shows quantitatively how the combined effect of crosstalk and misalignment contributes to decrease the signal-to-noise ratio.
In our experimental setup, error due to misalignment is negligible and crosstalk dominates, therefore $\sigma^2 + 4 w_0^2 \chi \simeq 4 w_0^2 \chi$.


\begin{figure}[t]
\centering
\includegraphics[width=0.9\linewidth]{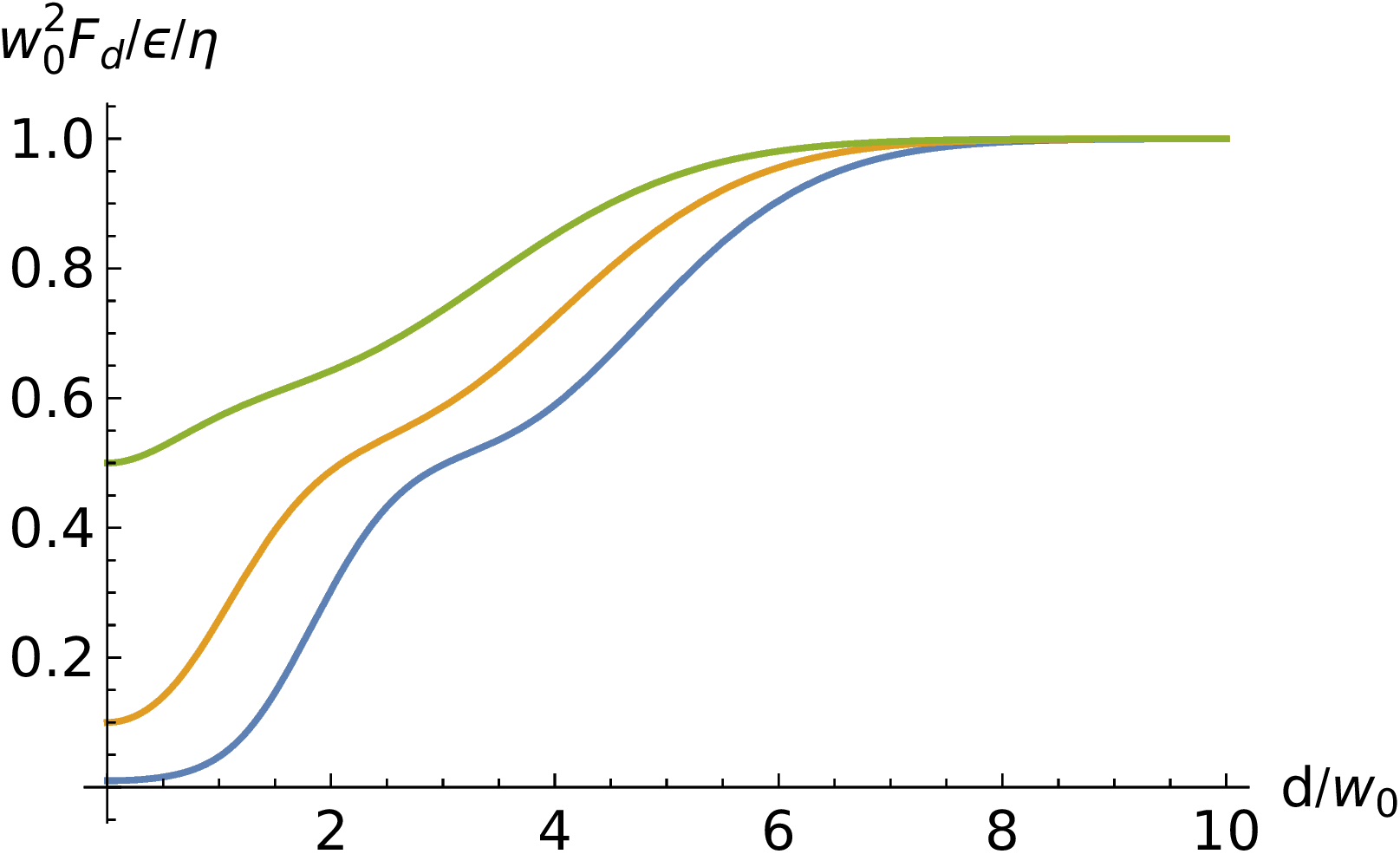}
\caption{Fisher information $F_d$ for the estimation of the source separation with DI, re-scaled by the factor $w_0^2/\epsilon/\eta$, plotted versus the dimensionless separation $d_a := d/w_0$.
From bottom to top, the curves are obtained for $\epsilon = 0.01,0.1,0.5$.
For small separation, the Fisher information is $F_d \simeq \epsilon^2/w_0^2$. For large separation, when the source are well separated, we have $F_d \simeq \epsilon/w_0^2$.}
\label{fi:Fd_d}
\end{figure}


\section{Comparison with direct imaging}\label{Sec:DI}

DI is represented by a measurement of the operators $n(x,y) = a^\dag (x,y) a(x,y)$ on the state in Eq.~(\ref{rho0}). 
For simplicity, we assume to know that the secondary source lays on the $x$ axis.
In the limit of infinite number of pixels (i.e.~infinite resolution), the probability density of detecting a photon at position $(x,y)$ in the image plane is
\begin{align}
p(x,y) & = \mathrm{Tr}[ n(x,y) \rho ] \\
& = \frac{\eta}{2\pi w_0^2} 
\left[ (1-\epsilon) 
  \,
    e^{-\frac{x^2}{2 w_0^2}}  
+ \epsilon 
  \,
    e^{-\frac{(x-d)^2}{2 w_0^2}} 
    \right]
        e^{-\frac{y^2}{2 w_0^2}}  
    \, .
\end{align}
The Fisher information associated to the estimation of the distance $d$ is 
\begin{align}
F_d = \int  \frac{dx dy}{p(x,y)} \left( \frac{\partial p(x,y)}{\partial d} \right)^2
= \int  \frac{dx}{p(x)} \left( \frac{\partial p(x)}{\partial d} \right)^2  \, ,
\end{align}
where $p(x)$ is the marginal probability. 
Note that since
\begin{align}
\frac{\partial p(x)}{\partial d} =
- \frac{\eta \epsilon}{\sqrt{2\pi w_0^2}} 
   \,
    \frac{x-d}{w_0^2} e^{-\frac{(x-d)^2}{2 w_0^2}} 
 \, ,
\end{align}
the Fisher information is proportional to $\epsilon^2$ if $d$ is comparable to or smaller than $w_0$. 
Otherwise, for $d \gg w_0$, when the two sources are well separated, it is proportional to $\epsilon$.
$F_d$ can be computed numerically.
Figure~\ref{fi:Fd_d} shows its dependence on the separation $d$ for different values of $\epsilon$. The plot shows that the Fisher information does not go to zero when $d$ approaches zero, but it goes to $\epsilon^2/w_0^2$. 
This is due to the fact that,  unlike previous works, here we align the optical system to the brighter source instead of the median point between the two sources (which coincides with the ``center of mass'' for equally bright sources).

Similarly, we can compute the Fisher information for the estimation of $\epsilon$,
\begin{align}
F_\epsilon = \int  \frac{dx dy}{p(x,y)} \left( \frac{\partial p(x,y}{\partial \epsilon} \right)^2
= \int  \frac{dx}{p(x)} \left( \frac{\partial p(x)}{\partial \epsilon} \right)^2  \, .
\end{align}
Since
\begin{align}
\frac{\partial p(x)}{\partial \epsilon} =
\frac{\eta}{\sqrt{2\pi w_0^2}} \left[
    e^{-\frac{(x-d)^2}{2 w_0^2}} 
    - e^{-\frac{x^2}{2 w_0^2}}
    \right]
 \, ,
\end{align}
$F_\epsilon$ is only mildly dependent on $\epsilon$.
By taking a Taylor expansion around $d_a := d/w_0 = 0$ we also see that $F_\epsilon \simeq \eta(d/w_0)^2$ in the sub-Rayleigh regime.

From the Fisher information, we can compute the ultimate precision in the estimation of these parameters from DI. We have
\begin{align}
\Delta_\text{DI} d \geq 1/\sqrt{n F_d } \, , 
\label{errdDI} \\
\Delta_\text{DI} \epsilon \geq 1/\sqrt{ n F_\epsilon } \, .
\label{erreDI}
\end{align}
These quantities can be compared with the estimates in Eqs.~(\ref{errd})-(\ref{erre}).
In the ideal case where noise is negligible, i.e.~for $\sigma^2 + 4 w_0^2 \chi \ll \epsilon d^2$, Eqs.~(\ref{errd})-(\ref{erre}) simplify to 
\begin{align}
\Delta d 
& = w_0 \sqrt{ \frac{1}{ \eta n \epsilon} } \, , 
\label{errd_} \\
\Delta \epsilon
& = \frac{2 w_0}{d} 
\sqrt{ \frac{ \epsilon }{\eta n}}
\, .
\label{erre_}
\end{align}
This shows a different scaling with respect to DI.
In particular, $\Delta d$ is proportional to $\epsilon^{-1/2}$ and independent of $d$, whereas $\Delta_\text{DI} d$ goes as $\epsilon^{-1}$ for small $d \ll w_0$.
Similarly, $\Delta \epsilon$ goes to zero as $\epsilon^{1/2}$ for small $\epsilon$, whereas $\Delta_\text{DI} \epsilon$ is nearly independent of $\epsilon$.

\begin{figure}[t]
\centering
\includegraphics[width=0.9\linewidth]{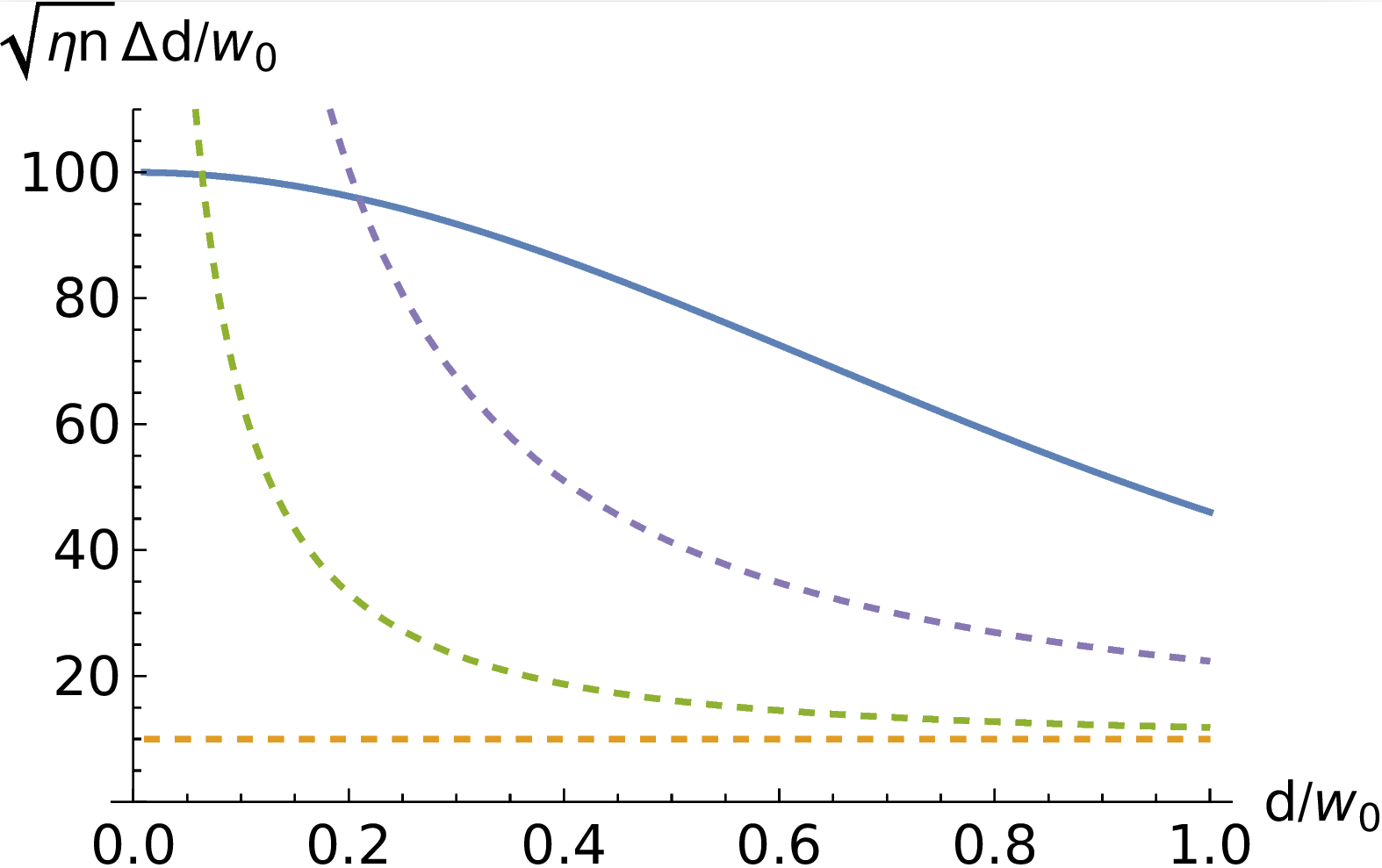}
\caption{Statistical error (re-scaled by a factor $\sqrt{\eta n}/w_0$) in the estimation of the separation $d$, plotted versus the dimensionless parameter $d_a = d/w_0$. 
For $\epsilon=0.01$, the solid line shows the error from noiseless DI, computed using Eq.~(\ref{errdDI}).
The dashed lines show the error for SPADE, in the presence of crosstalk, computed from Eq.~(\ref{errd}), for negligible alignment error ($\sigma=0$); from bottom to top the crosstalk error is $\chi=0, 0.001,0.01$.}
\label{fi:Delta_d}
\end{figure}

In conclusion, for small $\epsilon$ noiseless SPADE allows for a reduced error compared with DI, yielding an improvement by a factor $1/\sqrt{\epsilon}$. However, this improvement is limited by noise, in particular by crosstalk. 
As an example, Fig.~\ref{fi:Delta_d} shows the effect of crosstalk. The plot compares noiseless DI with noiseless and noisy SPADE. While noiseless SPADE outperforms noiseless DI by a factor up to $\sqrt{\epsilon}$, this advantage is quickly washed out by crosstalk. As the figure shows, a crosstalk factor $\chi$ of the order of $1\%$ already drastically reduces the performance of SPADE. 
As also discussed by Linowski at al.~\cite{ct2022}, in the presence of crosstalk SPADE outperforms DI only if the separation is not too small (note that our analysis of DI is different as we align the optical system to the brighter source and not to the median point).
However, our comparison is not entirely fair as we are considering ideal DI with an infinite number of arbitrary small pixels. In practice, DI would have finite number of pixels and would be as well affected by crosstalk between neighbor pixels.

\begin{figure}[t]
\centering
\includegraphics[width=0.9\linewidth]{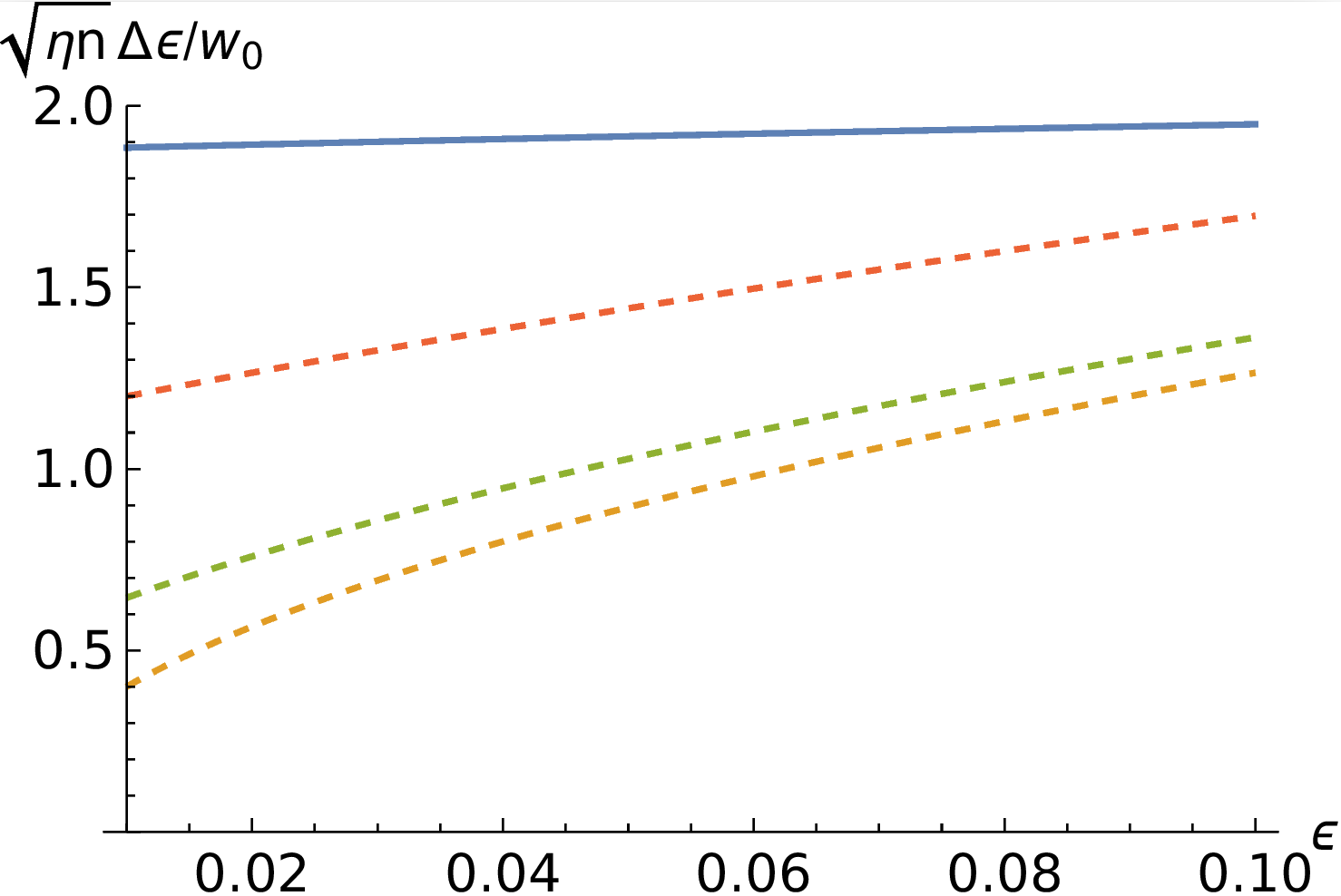}
\caption{Statistical error (re-scaled by a factor $\sqrt{\eta n}/w_0$) in the estimation of the relative intensity $\epsilon$, plotted versus $\epsilon$. 
For $d=0.5 w_0$, the solid line shows the error from noiseless DI, computed using Eq.~(\ref{erreDI}).
The dashed lines show the error for SPADE, in the presence of crosstalk, computed from Eq.~(\ref{erre}), for negligible alignment error ($\sigma=0$); from bottom to top the crosstalk error is $\chi=0, 0.001,0.005$.}
\label{fi:Delta_e}
\end{figure}

Similar conclusions can be drawn for the problem of estimating the parameter $\epsilon$; a comparison of the statistical error for noiseless DI and noiseless and noisy SPADE is shown in Fig.~\ref{fi:Delta_e}.


\section{Experimental setup}

\begin{figure}
\centering
\includegraphics[width=0.99\linewidth]{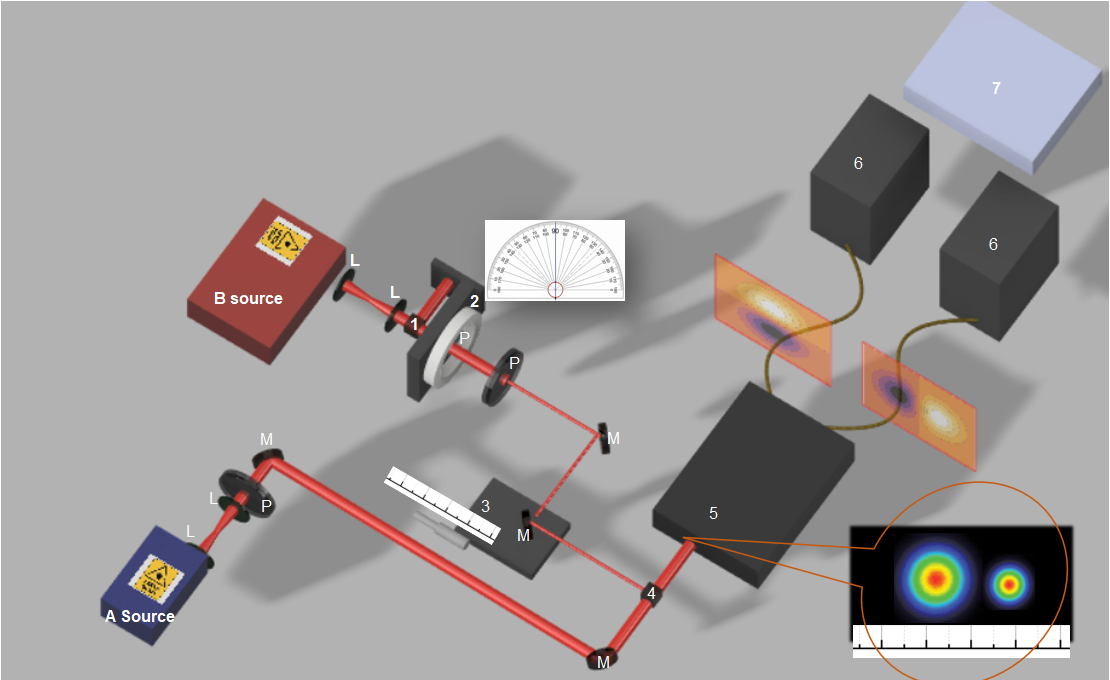}
\caption{\textbf{Experimental setup.} The weaker source (B source) is a heralded photon source that crosses a two-lens system (L) to match the beam waist with demultiplexer waist ($w_0=300$ $\mu m$) and then a polarizing beam splitter (1).
The transmitted component, used for the experiment, crosses a polarizer (P) mounted on a motorized rotation stage (2), a second polarizer (P), two mirrors (M), reflects on a beam splitter (4) and impinges on free-space input of the demultiplexer (5).
The brighter source (A source) crosses a two-lens (L) system to match the beam waist with demultiplexer waist ($300$ $\mu m$), a fixed polarizer (P), two mirrors (M), the beam splitter (4) where combines with B source and impinges on input port of demultiplexer(5).
The intensity and position of source B can be tuned using the motorized rotation stage (2) and translation stage (3).
The demultiplexer (5), PROTEUS-C from Cailabs, allows to perform intensity measurements on six HG mode, but just the $\text{HG}_{01}$ and  $\text{HG}_{10}$ modes are detected through two free-running InGaAs/InP single-photon avalanche diodes cooled at $-90^{\circ}$ (6), whose electrical output signal is detected and counted by a time tagger (7).}
\label{fi:setup}
\end{figure}

The experimental setup is shown in Fig.~\ref{fi:setup}. We combine two collimated sources (at telecom wavelength) on a non-polarizing beam splitter (NPBS) to mimic two point-like sources.
The brighter source is denoted as source A, the weaker one as source B, with intensity $I_A$ and $I_B$, respectively. The relative intensity is $I_B/I_A \simeq \epsilon$.


The source B is a heralded photon source. It consists of a $3 \, mW$ continuous-wave laser at $775 \, nm$ pumping a type II waveguide of a Periodically Poled Lithium Niobate (PPLN) crystal stabilized at $33.90^\circ$ (within $0.01^\circ$) by a proportional integrative derivative (PID) controller. The pump, by means of spontaneous parametric downconversion, generates about $500.000$ photon pairs (orthogonally polarized) per second at $1550 \, nm$ wavelength.
At the output, the beam crosses a two-lens system to match the beam waist with demultiplexer waist ($w_0=300 \, \mu m$) and a polarizer beam splitter, where only the transmitted photons are used for the experiment.
The other photons of the pairs, vertically polarized, are not measured. The intensity of the vertically polarized beam can be accurately tuned by means of two free-space polarizers. 
Note that, being the horizontal photon  not measured, the vertical one behaves as a thermal state.
In detail, the beam from the source B crosses a film polarizer mounted on a motorized rotation stage and then another film polarizer (at fixed angle).
In this way, by changing the angle of the rotation stage, is possible to tune the B beam intensity while maintaining the polarization fixed (being the second polarizer fixed).
We align the angle of the fixed polarizer with direction of maximum detector responsivity to the elctric field: in this way we are confident that the polarization of the B beam is fixed for any angle of the first polarizer bypassing the issue of polarization-dependency detector efficiency.
Finally, by means of two steering mirrors (the second being mounted on a micrometric translation stage to shift the beam position), the beam is coupled with free-space input port of the demultiplexer after reflection on a beam splitter used to overlap B and A beams.

The source A is a few $mW$ light emitting diode (LED) at $1550 \, nm$, attenuated by $90 \, dB$ using fiber attenuators, and collimated using
a two-lens system to match the beam waist with the demultiplexer waist.
As for the source B, a polarizer is used to align with the direction of maximum detector efficiency and a pair of steering mirrors are used to couple the beam with demultiplexer input after crossing a beam splitter where A and B beams are overlapped.
The position of the beam A remains fixed during the experiment, whereas the source B can be attenuated by changing the angle of rotation stage and shifted by moving the translator.
With this scheme it is possible to tune the separation $d$ and the intensity ratio $\epsilon$ independently.

The two beams are finally fed into a demultiplexer, PROTEUS-C model from Cailabs, which allows for intensity measurements on six HG modes. It accepts radiation from the free-space input port, and decomposes it in the lowest-order modes ($\text{HG}_{00}$, $\text{HG}_{01}$, $\text{HG}_{10}$, $\text{HG}_{11}$, $\text{HG}_{20}$, $\text{HG}_{02}$). The modes are coupled with six single-mode fibers following conversion into the $\text{HG}_{00}$ mode.
Finally, the modes $\text{HG}_{01}$ and $\text{HG}_{10}$ are coupled, through a single mode fiber, with free-running InGaAs/InP single-photon avalanche diodes cooled at $-90^{\circ} C$ and operating with dead time of $20 \, \mu s$ allowing a negligible dark count rate. The detectors generate electric pulse recorded by a time-to-digital converter. 
When the beams separation is zero (i.e., the beams overlap completely) and the overall source (A+B beam) has a circular symmetry, the power leaked into the $\text{HG}_{01}$ and $\text{HG}_{10}$ modes reaches its minimum value.

One of the main source of experimental error in the setup is due to crosstalk.
The crosstalk between $\text{HG}_{00}$ and $\text{HG}_{nm}$ may be quantified by the ratio 
$P_{nm} / P_{00}$, where $P_{nm}$ is the  power on the $\text{HG}_{nm}$ output channel if only $\text{HG}_{00}$ is injected with a power $P_{00}$ from the input.
The crosstalk is a limitation that is due to the presence of light in high order modes even if the incoming radiation is fully matched with the demultiplexer (in the ideal case of negligible crosstalk only $\text{HG}_{00}$ should be excited). 
The crosstalk factor $\chi$ introduced in Section~\ref{ssec:ct} is $\chi = P_{01} / P_{00} + P_{10} / P_{00}$.

In the experiment we operate the demultiplexer in the single-photon regime for the estimation of the beam intensity ratio $\epsilon = I_B/I_A$ and the beam separation $d$. The beams have to simulate two point-like sources with Gaussian PSF characterized by the demultiplexer waist $w_0$.
To simulate a situation where we do not know whether there is a single source or there are two sources, we align the system by maximizing the $\text{HG}_{00}$ output when the two sources are completely superimposed. 
Data are collected by translating the B beam and keeping the A beam centered.
In a practical scenario, one can only align the demultiplexer with the ``center of mass'' of light intensity, since the positions of the sources are unknown.
Consequently, the proper procedure for device calibration would be to maximize the power in $\text{HG}_{00}$ mode as we did, but then, to move both the beams in opposite directions. However, as discussed in Section~\ref{ssec:align} the two procedures are equivalent up to an error of order $\epsilon d$.


\subsection{Calibration and measurement}

We acquired the photon counts $C_{n}^{H_{01}}$ ($C_{n}^{H_{10}}$) in $\text{HG}_{01}$ ($\text{HG}_{10}$) modes. 
The total photon counts impinging on detectors coupled to $\text{HG}_{01}$ and $\text{HG}_{10}$ modes is then:
\begin{equation} \label{somma}
   n_1 =   C_n^{H_{01}}+C_n^{H_{10}} \, .
\end{equation}
We record $n_1$ by changing the sources separation $d$ and intensity ratio $\epsilon$ by acting on translation stage and rotation stage using LabView software.
The values of $d$ are changed between $-200 \, \mu m$ and $200 \, \mu m$ with a step size of $20 \, \mu m$ ($w_0=300 \, \mu m$). 


\begin{figure}[t]
\centering
\includegraphics[width=0.99\linewidth]{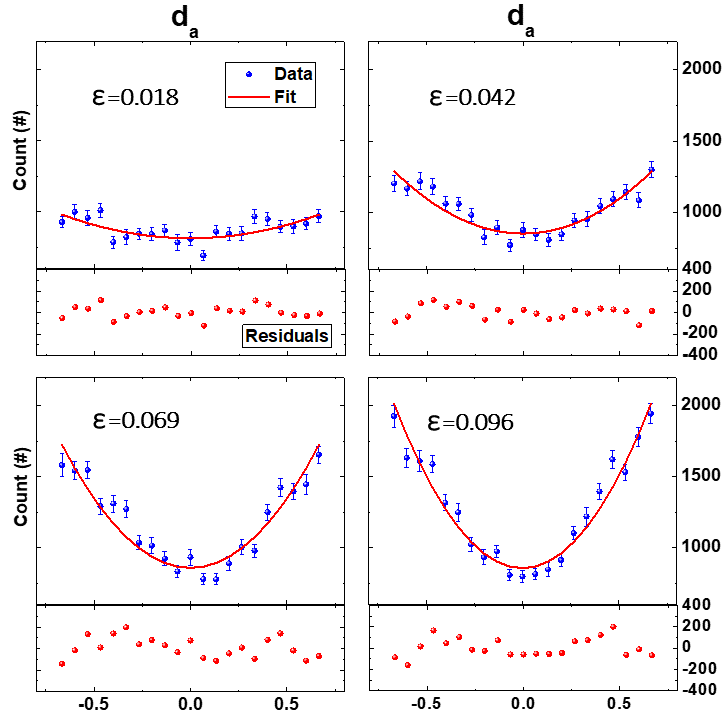}
\caption{Top panels: Counts $n_1$ as function of $d_a$, for four values of the intensity ratio $\epsilon$; the blue points are from experimental data, with error bars estimated upon repeated measurements; the red line shows the result of a quadratic regression on the data points.
Bottom panels: Residuals between experimental data and fitted curve. They appear random distributed indicating the quadratic function a good approximation of HG in the range of measurements.}
\label{fi:Fig1}
\end{figure}


\begin{figure}[t]
\centering
\includegraphics[width=0.99\linewidth]{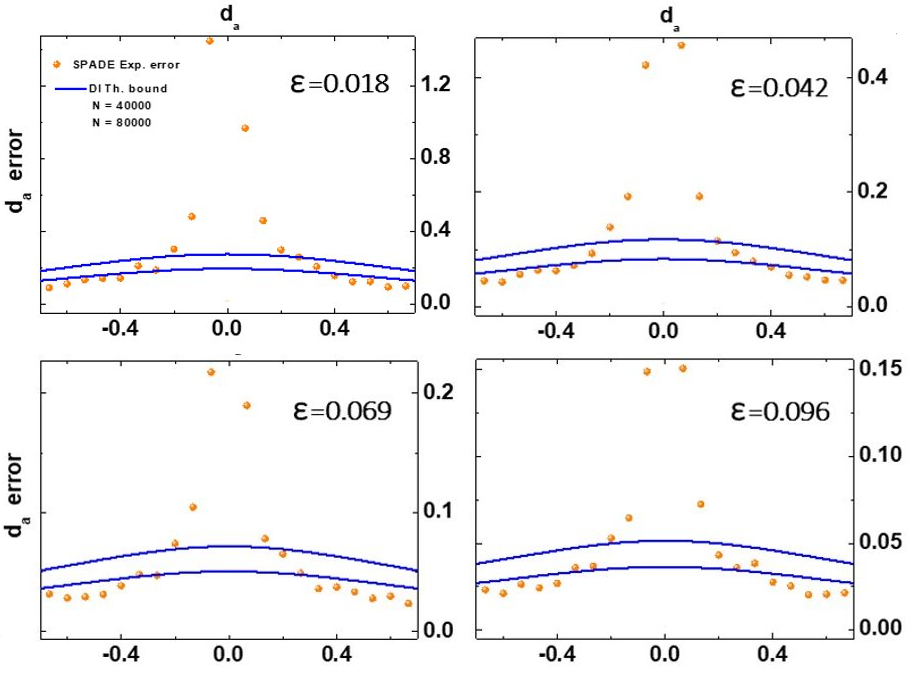}
\caption{Experimental uncertainty $\delta d_{a}$ (orange points) as function of $d_{a}$, for four different values of the intensity ratio $\epsilon$ and the theoretical error estimates (blue lines) for DI imaging calculated for $\eta n = 40000$ and $\eta n = 80000$ photons.}
\label{fi:Fig2}
\end{figure}


For each value of $\epsilon$ and $d$ we detected photons in $\text{HG}_{10} + \text{HG}_{01}$ modes during $1 \, s$ of acquisition time.
This is obtained by combining $N_m=100$ independent measures $n_1^i$, for $i=1,\dots, 100$, lasting $10 \, ms$ each.
Figure~\ref{fi:Fig1} shows the total counts $n_1(d_a)=\sum_{i=1}^{N_m} n_1^i(d_a)$ as function of the dimensionless separation $d_a := d/w_0$, for four different values of $\epsilon$ (blue points) and the corresponding quadratic regressions $n_1^F(d_a)=a_1+b_1 d_a^2$ (red line), where $a_1$ and $b_1$ (related to crosstalk and dark count) are fitting parameters. 
As discussed in Section~\ref{ssec:CRB}, a quadratic law is expected for small separations, $d_a \ll 1$.
%

We use the fitted curve $n_1^F(d_a)$ as calibration curve to measure the distance set on the translator stage.
We obtain the associated uncertainty $\delta d_a$ by error propagation from the uncertainty in the total photon count $n_1$,
\begin{align}
    \delta d_a = \left| \frac{\partial n_1^F}{\partial d_a} \right|^{-1} \delta n_1 \, ,
\end{align}
where $\delta n_1$ is the standard deviation of $n_1^i$ multiplied by~$\sqrt{N_m}$.

Figure~\ref{fi:Fig2} shows the experimental uncertainty $\delta d_a$ (orange points) compared with the theoretical bounds for ideal DI computed in Section~\ref{Sec:DI}. 
The latter are computed using the estimated photon number $\eta n$ during $1 \, s$ of acquisition time of, which is in the range between $4 \times 10^4$ and $8 \times 10^4$.
This shows that our setup, though affected by crosstalk and dark counts (errors due to residual misalignment are negligible), allows us to beat ideal and noiseless DI if $d_a$ is not too small.

We also note that our experimental errors are about 2-4 times larger than the SPADE theoretical bounds of Section \ref{ssec:ct} (computed using the experimental value $\chi 
= 0.0035$~\cite{Santamaria22}), and the gap increases at small separation $d_a \ll 1$.
The discrepancy is likely due to finite-sampling and noise in the mechanical mounts and imperfections of the translation stage. In fact, being the measurement performed by changing the position of translation stage, the imperfections of the translation generate increased uncertainty $\delta d_a$ that does not depend on the imaging system but just on the sources. In other words, in a real acquisition, these imperfections should not be present.

\begin{figure}[t]
\centering
\includegraphics[width=0.99\linewidth]{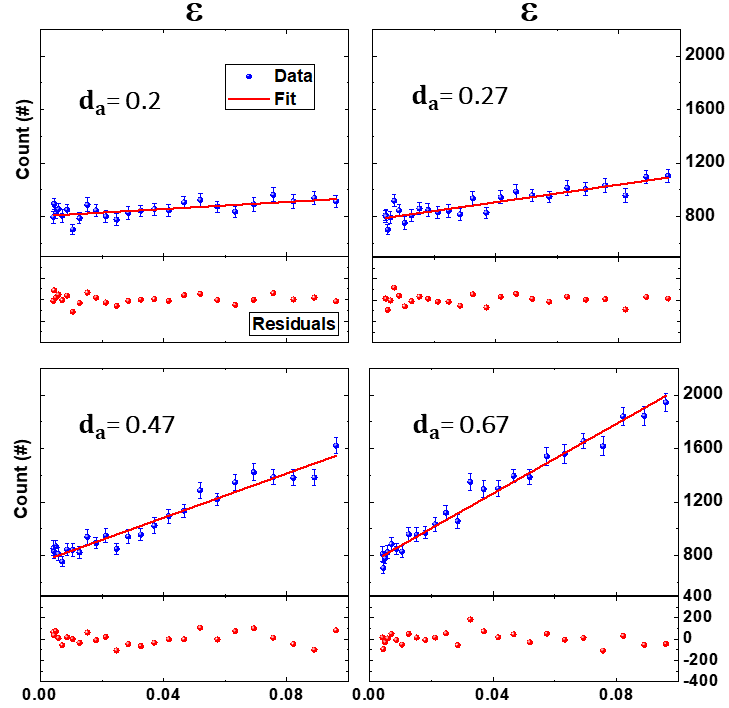}
\caption{Top panels: Counts $n_1$ as function of $\epsilon$ at four values of separation $d_a$ (blue points) with error bars obtained from $N_m$ repeated measurements; the red line show the result of a linear regression on the data points. Lower panels: residuals between experimental data and fitted curve. They appear random distributed confirming the proportionality between $\epsilon$ and $d_a$.}
\label{fi:Fig3}
\end{figure}

\begin{figure}[t]
\centering
\includegraphics[width=0.99\linewidth]{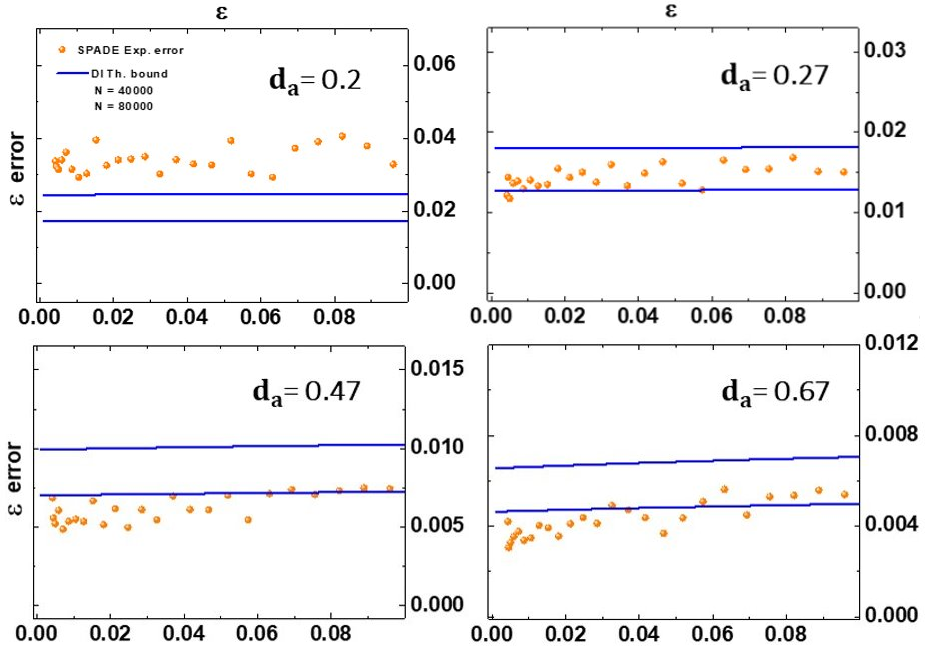}
\caption{Experimental uncertainty $\delta \epsilon$ (orange points) as function of $\epsilon$ at four different separation $d_a$ and corresponding
theoretical error estimates (blue lines) for DI imaging calculated for $\eta n = 40000$ and $\eta n = 80000$  photons.}
\label{fi:Fig4}
\end{figure}

We carry out analogous calibration and measurement for the estimation of the parameter $\epsilon$.
In Fig.~\ref{fi:Fig3} we represent the counts $n_1$ (blue points) as function of $\epsilon$ ratio at fixed distance, and the result of a linear regression (red line) that we use as calibration curve.
The experimental uncertainty $\delta \epsilon$ is obtained as 
\begin{align}
    \delta \epsilon = \left| \frac{\partial n_1^F}{\partial \epsilon} \right|^{-1} \delta n_1 \, .
\end{align}
In Fig.~\ref{fi:Fig4} the experimental uncertainty $\delta \epsilon$ is compared with theoretical bounds for ideal DI. As one would expect, SPADE performs better than DI for larger values of $d_a$ and small values of $\epsilon$.
Also, the experimental errors are larger than the SPADE theoretical error bounds (less than a factor two).
In this case the discrepancy is smaller compared to the discrepancy on the measure of $d_a$, most likely because the measurements are performed by changing $\epsilon$ without touching the translation stage.


\section{Conclusions}

In a diffraction-limited optical system, the resolution of direct imaging (DI) is limited by the width of the point-spread function, according to the Rayleigh criterion.
However, when interferometric measurements are employed, the information carried by the phase of the field may allow us to beat the Rayleigh resolution limit.
This is particularly important in the regime of single-photon imaging, where interferometric measurements achieve the ultimate precision limit as set by the quantum Cramér-Rao bound.

In this work we have demonstrated single-photon spatial-mode demultiplexing (SPADE) by exploiting multi-plane light conversion technology, to sort the transverse field into its Hermite-Gaussian (HG) components. 
Working with two single-photon avalanche diodes, we have detected the lowest-order modes $\text{HG}_{01}$ and $\text{HG}_{10}$, which are sufficient to estimate either the separation between two point-like sources (if the relative intensity is known), or their relative intensity $\epsilon$ (if the separation is known).
Unlike previous experimental works, here we have focused on sources of unequal intensity.
This is motivated by the fact that pairs of natural sources typically have unequal intensities. Of particular interest is the situation when the primary source is much brighter than the secondary one, $\epsilon \ll 1$, which suggests application to exoplanet imaging and detection.
In fact, it is in this regime that SPADE may outperform DI by a factor $\epsilon^{-1/2}$.

Our experimental errors are dominated by crosstalk and finite-sampling fluctuations are about three times larger than the asymptotic theoretical estimates.
Nevertheless, we have been able to demonstrate an error in parameter estimation below what could be achieved by ideal, noiseless DI.
By using larger samples and suitable opto-mechanical mounts we may substantially reduce the discrepancy with the theoretical bounds, potentially allowing SPADE to beat noiseless DI by a greater extent and for smaller values of the source separation.
In particular, the advantage of SPADE of DI may be more easily observed for smaller value of $\epsilon$; whereas in our experiment the factor $\epsilon^{-1/2}$ was always smaller than $10$.
Moreover, there is room for experimental improvement by using faster and more efficient detectors (e.g.~nanowires).
Finally, the crucial parameter is the crosstalk value; this is not a fundamental limitation and may be improved by future advances in manufacturing demultiplexer devices.

\vspace{0.5cm}

We acknowledge support from:
Italian Space Agency – Subdiffraction Quantum Imaging (SQI) n.~2023-13-HH.0;
MOlecular spectroscopy for Space science and quantum physics Test (MOST); and
European Union – Next Generation EU: PNRR MUR project PE0000023-NQSTI.
We thank Cailabs, 38 boulevard Albert 1er, 35200 Rennes, France.


\begin{thebibliography}{39}
\expandafter\ifx\csname natexlab\endcsname\relax\def\natexlab#1{#1}\fi
\expandafter\ifx\csname bibnamefont\endcsname\relax
  \def\bibnamefont#1{#1}\fi
\expandafter\ifx\csname bibfnamefont\endcsname\relax
  \def\bibfnamefont#1{#1}\fi
\expandafter\ifx\csname citenamefont\endcsname\relax
  \def\citenamefont#1{#1}\fi
\expandafter\ifx\csname url\endcsname\relax
  \def\url#1{\texttt{#1}}\fi
\expandafter\ifx\csname urlprefix\endcsname\relax\def\urlprefix{URL }\fi
\providecommand{\bibinfo}[2]{#2}
\providecommand{\eprint}[2][]{\url{#2}}

\bibitem[{\citenamefont{Rayleigh}(1879)}]{Rayleigh1879}
\bibinfo{author}{\bibfnamefont{L.}~\bibnamefont{Rayleigh}},
  \bibinfo{journal}{The London, Edinburgh, and Dublin Philosophical Magazine
  and Journal of Science} \textbf{\bibinfo{volume}{8}}, \bibinfo{pages}{261}
  (\bibinfo{year}{1879}),
  \urlprefix\url{https://doi.org/10.1080/14786447908639684}.

\bibitem[{\citenamefont{Goodman}(2008)}]{goodman2008introduction}
\bibinfo{author}{\bibfnamefont{J.}~\bibnamefont{Goodman}},
  \emph{\bibinfo{title}{Introduction to Fourier optics}}
  (\bibinfo{publisher}{McGraw-hill}, \bibinfo{year}{2008}).

\bibitem[{\citenamefont{Hell and Wichmann}(1994)}]{Hell1994}
\bibinfo{author}{\bibfnamefont{S.~W.} \bibnamefont{Hell}} \bibnamefont{and}
  \bibinfo{author}{\bibfnamefont{J.}~\bibnamefont{Wichmann}},
  \bibinfo{journal}{Optics Letters} \textbf{\bibinfo{volume}{19}},
  \bibinfo{pages}{780} (\bibinfo{year}{1994}),
  \urlprefix\url{https://doi.org/10.1364/ol.19.000780}.

\bibitem[{\citenamefont{Kolobov}(2007)}]{kolobov2007quantum}
\bibinfo{author}{\bibfnamefont{M.~I.} \bibnamefont{Kolobov}},
  \emph{\bibinfo{title}{Quantum imaging}} (\bibinfo{publisher}{Springer Science
  \& Business Media}, \bibinfo{year}{2007}).

\bibitem[{\citenamefont{Boto et~al.}(2000)\citenamefont{Boto, Kok, Abrams,
  Braunstein, Williams, and Dowling}}]{JPD}
\bibinfo{author}{\bibfnamefont{A.~N.} \bibnamefont{Boto}},
  \bibinfo{author}{\bibfnamefont{P.}~\bibnamefont{Kok}},
  \bibinfo{author}{\bibfnamefont{D.~S.} \bibnamefont{Abrams}},
  \bibinfo{author}{\bibfnamefont{S.~L.} \bibnamefont{Braunstein}},
  \bibinfo{author}{\bibfnamefont{C.~P.} \bibnamefont{Williams}},
  \bibnamefont{and} \bibinfo{author}{\bibfnamefont{J.~P.}
  \bibnamefont{Dowling}}, \bibinfo{journal}{Phys. Rev. Lett.}
  \textbf{\bibinfo{volume}{85}}, \bibinfo{pages}{2733} (\bibinfo{year}{2000}),
  \urlprefix\url{https://link.aps.org/doi/10.1103/PhysRevLett.85.2733}.

\bibitem[{\citenamefont{Giovannetti et~al.}(2009)\citenamefont{Giovannetti,
  Lloyd, Maccone, and Shapiro}}]{Giovannetti}
\bibinfo{author}{\bibfnamefont{V.}~\bibnamefont{Giovannetti}},
  \bibinfo{author}{\bibfnamefont{S.}~\bibnamefont{Lloyd}},
  \bibinfo{author}{\bibfnamefont{L.}~\bibnamefont{Maccone}}, \bibnamefont{and}
  \bibinfo{author}{\bibfnamefont{J.~H.} \bibnamefont{Shapiro}},
  \bibinfo{journal}{Phys. Rev. A} \textbf{\bibinfo{volume}{79}},
  \bibinfo{pages}{013827} (\bibinfo{year}{2009}),
  \urlprefix\url{https://link.aps.org/doi/10.1103/PhysRevA.79.013827}.

\bibitem[{\citenamefont{Helstrom}(1976)}]{Helstrom_book}
\bibinfo{author}{\bibfnamefont{C.~W.} \bibnamefont{Helstrom}},
  \emph{\bibinfo{title}{Quantum Detection and Estimation Theory}}
  (\bibinfo{publisher}{Academic Press, New York}, \bibinfo{year}{1976}).

\bibitem[{\citenamefont{Tsang et~al.}(2016)\citenamefont{Tsang, Nair, and
  Lu}}]{Tsang2016}
\bibinfo{author}{\bibfnamefont{M.}~\bibnamefont{Tsang}},
  \bibinfo{author}{\bibfnamefont{R.}~\bibnamefont{Nair}}, \bibnamefont{and}
  \bibinfo{author}{\bibfnamefont{X.-M.} \bibnamefont{Lu}},
  \bibinfo{journal}{Physical Review X} \textbf{\bibinfo{volume}{6}},
  \bibinfo{pages}{031033} (\bibinfo{year}{2016}),
  \urlprefix\url{https://doi.org/10.1103/physrevx.6.031033}.

\bibitem[{\citenamefont{Paris}(2009)}]{Paris}
\bibinfo{author}{\bibfnamefont{M.~G.~A.} \bibnamefont{Paris}},
  \bibinfo{journal}{International Journal of Quantum Information}
  \textbf{\bibinfo{volume}{07}}, \bibinfo{pages}{125} (\bibinfo{year}{2009}),
  \urlprefix\url{https://doi.org/10.1142/S0219749909004839}.

\bibitem[{\citenamefont{Sidhu and Kok}(2020)}]{Sidhu}
\bibinfo{author}{\bibfnamefont{J.~S.} \bibnamefont{Sidhu}} \bibnamefont{and}
  \bibinfo{author}{\bibfnamefont{P.}~\bibnamefont{Kok}}, \bibinfo{journal}{AVS
  Quantum Science} \textbf{\bibinfo{volume}{2}}, \bibinfo{pages}{014701}
  (\bibinfo{year}{2020}), ISSN \bibinfo{issn}{2639-0213},
  \urlprefix\url{https://doi.org/10.1116/1.5119961}.

\bibitem[{\citenamefont{Nair and Tsang}(2016)}]{Nair:16}
\bibinfo{author}{\bibfnamefont{R.}~\bibnamefont{Nair}} \bibnamefont{and}
  \bibinfo{author}{\bibfnamefont{M.}~\bibnamefont{Tsang}},
  \bibinfo{journal}{Opt. Express} \textbf{\bibinfo{volume}{24}},
  \bibinfo{pages}{3684} (\bibinfo{year}{2016}),
  \urlprefix\url{https://opg.optica.org/oe/abstract.cfm?URI=oe-24-4-3684}.

\bibitem[{\citenamefont{Dutton et~al.}(2019)\citenamefont{Dutton, Kerviche,
  Ashok, and Guha}}]{Dutton}
\bibinfo{author}{\bibfnamefont{Z.}~\bibnamefont{Dutton}},
  \bibinfo{author}{\bibfnamefont{R.}~\bibnamefont{Kerviche}},
  \bibinfo{author}{\bibfnamefont{A.}~\bibnamefont{Ashok}}, \bibnamefont{and}
  \bibinfo{author}{\bibfnamefont{S.}~\bibnamefont{Guha}},
  \bibinfo{journal}{Phys. Rev. A} \textbf{\bibinfo{volume}{99}},
  \bibinfo{pages}{033847} (\bibinfo{year}{2019}),
  \urlprefix\url{https://link.aps.org/doi/10.1103/PhysRevA.99.033847}.

\bibitem[{\citenamefont{Lupo et~al.}(2020)\citenamefont{Lupo, Huang, and
  Kok}}]{PRL2020}
\bibinfo{author}{\bibfnamefont{C.}~\bibnamefont{Lupo}},
  \bibinfo{author}{\bibfnamefont{Z.}~\bibnamefont{Huang}}, \bibnamefont{and}
  \bibinfo{author}{\bibfnamefont{P.}~\bibnamefont{Kok}},
  \bibinfo{journal}{Phys. Rev. Lett.} \textbf{\bibinfo{volume}{124}},
  \bibinfo{pages}{080503} (\bibinfo{year}{2020}),
  \urlprefix\url{https://link.aps.org/doi/10.1103/PhysRevLett.124.080503}.

\bibitem[{\citenamefont{Lu et~al.}(2018)\citenamefont{Lu, Krovi, Nair, Guha,
  and Shapiro}}]{lu2018quantum}
\bibinfo{author}{\bibfnamefont{X.-M.} \bibnamefont{Lu}},
  \bibinfo{author}{\bibfnamefont{H.}~\bibnamefont{Krovi}},
  \bibinfo{author}{\bibfnamefont{R.}~\bibnamefont{Nair}},
  \bibinfo{author}{\bibfnamefont{S.}~\bibnamefont{Guha}}, \bibnamefont{and}
  \bibinfo{author}{\bibfnamefont{J.~H.} \bibnamefont{Shapiro}},
  \bibinfo{journal}{npj Quantum Information} \textbf{\bibinfo{volume}{4}},
  \bibinfo{pages}{1} (\bibinfo{year}{2018}).

\bibitem[{\citenamefont{Huang and Lupo}(2021)}]{PRL2021}
\bibinfo{author}{\bibfnamefont{Z.}~\bibnamefont{Huang}} \bibnamefont{and}
  \bibinfo{author}{\bibfnamefont{C.}~\bibnamefont{Lupo}},
  \bibinfo{journal}{Phys. Rev. Lett.} \textbf{\bibinfo{volume}{127}},
  \bibinfo{pages}{130502} (\bibinfo{year}{2021}),
  \urlprefix\url{https://doi.org/10.1103/PhysRevLett.127.130502}.

\bibitem[{\citenamefont{Schlichtholz et~al.}(2023)\citenamefont{Schlichtholz,
  Linowski, Walschaers, Treps, Łukasz Rudnicki, and Sorelli}}]{Sorelli}
\bibinfo{author}{\bibfnamefont{K.}~\bibnamefont{Schlichtholz}},
  \bibinfo{author}{\bibfnamefont{T.}~\bibnamefont{Linowski}},
  \bibinfo{author}{\bibfnamefont{M.}~\bibnamefont{Walschaers}},
  \bibinfo{author}{\bibfnamefont{N.}~\bibnamefont{Treps}},
  \bibinfo{author}{\bibnamefont{Łukasz Rudnicki}}, \bibnamefont{and}
  \bibinfo{author}{\bibfnamefont{G.}~\bibnamefont{Sorelli}},
  \bibinfo{journal}{arXiv preprint arXiv:2303.02654}  (\bibinfo{year}{2023}),
  \urlprefix\url{https://arxiv.org/abs/2303.02654}.

\bibitem[{\citenamefont{Grace and Guha}(2022)}]{GraceGuha}
\bibinfo{author}{\bibfnamefont{M.~R.} \bibnamefont{Grace}} \bibnamefont{and}
  \bibinfo{author}{\bibfnamefont{S.}~\bibnamefont{Guha}},
  \bibinfo{journal}{Phys. Rev. Lett.} \textbf{\bibinfo{volume}{129}},
  \bibinfo{pages}{180502} (\bibinfo{year}{2022}),
  \urlprefix\url{https://link.aps.org/doi/10.1103/PhysRevLett.129.180502}.

\bibitem[{\citenamefont{Tsang}(2019{\natexlab{a}})}]{Trev}
\bibinfo{author}{\bibfnamefont{M.}~\bibnamefont{Tsang}},
  \bibinfo{journal}{Contemporary Physics} \textbf{\bibinfo{volume}{60}},
  \bibinfo{pages}{279} (\bibinfo{year}{2019}{\natexlab{a}}),
  \urlprefix\url{https://doi.org/10.1080/00107514.2020.1736375}.

\bibitem[{\citenamefont{Pa{\'{u}}r et~al.}(2016)\citenamefont{Pa{\'{u}}r,
  Stoklasa, Hradil, S{\'{a}}nchez-Soto, and Rehacek}}]{Par2016}
\bibinfo{author}{\bibfnamefont{M.}~\bibnamefont{Pa{\'{u}}r}},
  \bibinfo{author}{\bibfnamefont{B.}~\bibnamefont{Stoklasa}},
  \bibinfo{author}{\bibfnamefont{Z.}~\bibnamefont{Hradil}},
  \bibinfo{author}{\bibfnamefont{L.~L.} \bibnamefont{S{\'{a}}nchez-Soto}},
  \bibnamefont{and} \bibinfo{author}{\bibfnamefont{J.}~\bibnamefont{Rehacek}},
  \bibinfo{journal}{Optica} \textbf{\bibinfo{volume}{3}}, \bibinfo{pages}{1144}
  (\bibinfo{year}{2016}),
  \urlprefix\url{https://doi.org/10.1364/optica.3.001144}.

\bibitem[{\citenamefont{Pa\'{u}r et~al.}(2018)\citenamefont{Pa\'{u}r, Stoklasa,
  Grover, Krzic, S\'{a}nchez-Soto, Hradil, and \v{R}eh\'{a}\v{c}ek}}]{Paur:18}
\bibinfo{author}{\bibfnamefont{M.}~\bibnamefont{Pa\'{u}r}},
  \bibinfo{author}{\bibfnamefont{B.}~\bibnamefont{Stoklasa}},
  \bibinfo{author}{\bibfnamefont{J.}~\bibnamefont{Grover}},
  \bibinfo{author}{\bibfnamefont{A.}~\bibnamefont{Krzic}},
  \bibinfo{author}{\bibfnamefont{L.~L.} \bibnamefont{S\'{a}nchez-Soto}},
  \bibinfo{author}{\bibfnamefont{Z.}~\bibnamefont{Hradil}}, \bibnamefont{and}
  \bibinfo{author}{\bibfnamefont{J.}~\bibnamefont{\v{R}eh\'{a}\v{c}ek}},
  \bibinfo{journal}{Optica} \textbf{\bibinfo{volume}{5}}, \bibinfo{pages}{1177}
  (\bibinfo{year}{2018}),
  \urlprefix\url{https://opg.optica.org/optica/abstract.cfm?URI=optica-5-10-1177}.

\bibitem[{\citenamefont{Zhou et~al.}(2019)\citenamefont{Zhou, Yang, Hassett,
  Rafsanjani, Mirhosseini, Vamivakas, Jordan, Shi, and Boyd}}]{Zhou2019}
\bibinfo{author}{\bibfnamefont{Y.}~\bibnamefont{Zhou}},
  \bibinfo{author}{\bibfnamefont{J.}~\bibnamefont{Yang}},
  \bibinfo{author}{\bibfnamefont{J.~D.} \bibnamefont{Hassett}},
  \bibinfo{author}{\bibfnamefont{S.~M.~H.} \bibnamefont{Rafsanjani}},
  \bibinfo{author}{\bibfnamefont{M.}~\bibnamefont{Mirhosseini}},
  \bibinfo{author}{\bibfnamefont{A.~N.} \bibnamefont{Vamivakas}},
  \bibinfo{author}{\bibfnamefont{A.~N.} \bibnamefont{Jordan}},
  \bibinfo{author}{\bibfnamefont{Z.}~\bibnamefont{Shi}}, \bibnamefont{and}
  \bibinfo{author}{\bibfnamefont{R.~W.} \bibnamefont{Boyd}},
  \bibinfo{journal}{Optica} \textbf{\bibinfo{volume}{6}}, \bibinfo{pages}{534}
  (\bibinfo{year}{2019}),
  \urlprefix\url{https://doi.org/10.1364/optica.6.000534}.

\bibitem[{\citenamefont{Zhou et~al.}(2023)\citenamefont{Zhou, Xin, Li, and
  Lu}}]{Zhou:23}
\bibinfo{author}{\bibfnamefont{C.}~\bibnamefont{Zhou}},
  \bibinfo{author}{\bibfnamefont{J.}~\bibnamefont{Xin}},
  \bibinfo{author}{\bibfnamefont{Y.}~\bibnamefont{Li}}, \bibnamefont{and}
  \bibinfo{author}{\bibfnamefont{X.-M.} \bibnamefont{Lu}},
  \bibinfo{journal}{Opt. Express} \textbf{\bibinfo{volume}{31}},
  \bibinfo{pages}{19336} (\bibinfo{year}{2023}),
  \urlprefix\url{https://opg.optica.org/oe/abstract.cfm?URI=oe-31-12-19336}.

\bibitem[{\citenamefont{Tham et~al.}(2017)\citenamefont{Tham, Ferretti, and
  Steinberg}}]{Tham2017}
\bibinfo{author}{\bibfnamefont{W.-K.} \bibnamefont{Tham}},
  \bibinfo{author}{\bibfnamefont{H.}~\bibnamefont{Ferretti}}, \bibnamefont{and}
  \bibinfo{author}{\bibfnamefont{A.~M.} \bibnamefont{Steinberg}},
  \bibinfo{journal}{Physical Review Letters} \textbf{\bibinfo{volume}{118}},
  \bibinfo{pages}{070801} (\bibinfo{year}{2017}),
  \urlprefix\url{https://doi.org/10.1103/physrevlett.118.070801}.

\bibitem[{\citenamefont{Salit et~al.}(2020)\citenamefont{Salit, Klein, and
  Lust}}]{Salit:20}
\bibinfo{author}{\bibfnamefont{M.}~\bibnamefont{Salit}},
  \bibinfo{author}{\bibfnamefont{J.}~\bibnamefont{Klein}}, \bibnamefont{and}
  \bibinfo{author}{\bibfnamefont{L.}~\bibnamefont{Lust}},
  \bibinfo{journal}{Appl. Opt.} \textbf{\bibinfo{volume}{59}},
  \bibinfo{pages}{5319} (\bibinfo{year}{2020}),
  \urlprefix\url{https://opg.optica.org/ao/abstract.cfm?URI=ao-59-17-5319}.

\bibitem[{\citenamefont{Boucher et~al.}(2020)\citenamefont{Boucher, Fabre,
  Labroille, and Treps}}]{Boucher2020}
\bibinfo{author}{\bibfnamefont{P.}~\bibnamefont{Boucher}},
  \bibinfo{author}{\bibfnamefont{C.}~\bibnamefont{Fabre}},
  \bibinfo{author}{\bibfnamefont{G.}~\bibnamefont{Labroille}},
  \bibnamefont{and} \bibinfo{author}{\bibfnamefont{N.}~\bibnamefont{Treps}},
  \bibinfo{journal}{Optica} \textbf{\bibinfo{volume}{7}}, \bibinfo{pages}{1621}
  (\bibinfo{year}{2020}),
  \urlprefix\url{https://doi.org/10.1364/optica.404746}.

\bibitem[{\citenamefont{Santamaria et~al.}(2022)\citenamefont{Santamaria,
  Pallotti, de~Cumis, Dequal, and Lupo}}]{Santamaria22}
\bibinfo{author}{\bibfnamefont{L.}~\bibnamefont{Santamaria}},
  \bibinfo{author}{\bibfnamefont{D.}~\bibnamefont{Pallotti}},
  \bibinfo{author}{\bibfnamefont{M.~S.} \bibnamefont{de~Cumis}},
  \bibinfo{author}{\bibfnamefont{D.}~\bibnamefont{Dequal}}, \bibnamefont{and}
  \bibinfo{author}{\bibfnamefont{C.}~\bibnamefont{Lupo}},
  \bibinfo{journal}{arXiv preprint arXiv:2206.05246}  (\bibinfo{year}{2022}),
  \urlprefix\url{https://arxiv.org/abs/2206.05246}.

\bibitem[{\citenamefont{Rouvière et~al.}(2022)\citenamefont{Rouvière, Barral,
  Grateau, Karuseichyk, Sorelli, Walschaers, and Treps}}]{Treps23}
\bibinfo{author}{\bibfnamefont{C.}~\bibnamefont{Rouvière}},
  \bibinfo{author}{\bibfnamefont{D.}~\bibnamefont{Barral}},
  \bibinfo{author}{\bibfnamefont{A.}~\bibnamefont{Grateau}},
  \bibinfo{author}{\bibfnamefont{I.}~\bibnamefont{Karuseichyk}},
  \bibinfo{author}{\bibfnamefont{G.}~\bibnamefont{Sorelli}},
  \bibinfo{author}{\bibfnamefont{M.}~\bibnamefont{Walschaers}},
  \bibnamefont{and} \bibinfo{author}{\bibfnamefont{N.}~\bibnamefont{Treps}},
  \bibinfo{journal}{arXiv preprint arXiv:2306.11916}  (\bibinfo{year}{2022}),
  \urlprefix\url{https://arxiv.org/abs/2306.11916}.

\bibitem[{\citenamefont{Pushkina et~al.}(2021)\citenamefont{Pushkina, Maltese,
  Costa-Filho, Patel, and Lvovsky}}]{Lvovsky}
\bibinfo{author}{\bibfnamefont{A.~A.} \bibnamefont{Pushkina}},
  \bibinfo{author}{\bibfnamefont{G.}~\bibnamefont{Maltese}},
  \bibinfo{author}{\bibfnamefont{J.~I.} \bibnamefont{Costa-Filho}},
  \bibinfo{author}{\bibfnamefont{P.}~\bibnamefont{Patel}}, \bibnamefont{and}
  \bibinfo{author}{\bibfnamefont{A.~I.} \bibnamefont{Lvovsky}},
  \bibinfo{journal}{Phys. Rev. Lett.} \textbf{\bibinfo{volume}{127}},
  \bibinfo{pages}{253602} (\bibinfo{year}{2021}),
  \urlprefix\url{https://link.aps.org/doi/10.1103/PhysRevLett.127.253602}.

\bibitem[{\citenamefont{Frank et~al.}(2022)\citenamefont{Frank, Duplinskiy,
  Bearne, and Lvovsky}}]{Lvovsky2023}
\bibinfo{author}{\bibfnamefont{J.}~\bibnamefont{Frank}},
  \bibinfo{author}{\bibfnamefont{A.}~\bibnamefont{Duplinskiy}},
  \bibinfo{author}{\bibfnamefont{K.}~\bibnamefont{Bearne}}, \bibnamefont{and}
  \bibinfo{author}{\bibfnamefont{A.~I.} \bibnamefont{Lvovsky}},
  \bibinfo{journal}{arXiv preprint arXiv:2304.09773}  (\bibinfo{year}{2022}),
  \urlprefix\url{https://arxiv.org/abs/2304.09773}.

\bibitem[{\citenamefont{Gessner et~al.}(2020)\citenamefont{Gessner, Fabre, and
  Treps}}]{Gessner2020}
\bibinfo{author}{\bibfnamefont{M.}~\bibnamefont{Gessner}},
  \bibinfo{author}{\bibfnamefont{C.}~\bibnamefont{Fabre}}, \bibnamefont{and}
  \bibinfo{author}{\bibfnamefont{N.}~\bibnamefont{Treps}},
  \bibinfo{journal}{Physical Review Letters} \textbf{\bibinfo{volume}{125}},
  \bibinfo{pages}{100501} (\bibinfo{year}{2020}),
  \urlprefix\url{https://doi.org/10.1103/physrevlett.125.100501}.

\bibitem[{\citenamefont{Len et~al.}(2020)\citenamefont{Len, Datta, Parniak, and
  Banaszek}}]{Banaszek}
\bibinfo{author}{\bibfnamefont{Y.~L.} \bibnamefont{Len}},
  \bibinfo{author}{\bibfnamefont{C.}~\bibnamefont{Datta}},
  \bibinfo{author}{\bibfnamefont{M.}~\bibnamefont{Parniak}}, \bibnamefont{and}
  \bibinfo{author}{\bibfnamefont{K.}~\bibnamefont{Banaszek}},
  \bibinfo{journal}{International Journal of Quantum Information}
  \textbf{\bibinfo{volume}{18}}, \bibinfo{pages}{1941015}
  (\bibinfo{year}{2020}), 
  \urlprefix\url{https://doi.org/10.1142/S0219749919410156}.

\bibitem[{\citenamefont{Lupo}(2020)}]{PhysRevA.101.022323}
\bibinfo{author}{\bibfnamefont{C.}~\bibnamefont{Lupo}}, \bibinfo{journal}{Phys.
  Rev. A} \textbf{\bibinfo{volume}{101}}, \bibinfo{pages}{022323}
  (\bibinfo{year}{2020}),
  \urlprefix\url{https://link.aps.org/doi/10.1103/PhysRevA.101.022323}.

\bibitem[{\citenamefont{Linowski et~al.}(2022)\citenamefont{Linowski,
  Schlichtholz, Sorelli, Gessner, Walschaers, Treps, and Łukasz
  Rudnicki}}]{ct2022}
\bibinfo{author}{\bibfnamefont{T.}~\bibnamefont{Linowski}},
  \bibinfo{author}{\bibfnamefont{K.}~\bibnamefont{Schlichtholz}},
  \bibinfo{author}{\bibfnamefont{G.}~\bibnamefont{Sorelli}},
  \bibinfo{author}{\bibfnamefont{M.}~\bibnamefont{Gessner}},
  \bibinfo{author}{\bibfnamefont{M.}~\bibnamefont{Walschaers}},
  \bibinfo{author}{\bibfnamefont{N.}~\bibnamefont{Treps}}, \bibnamefont{and}
  \bibinfo{author}{\bibnamefont{Łukasz Rudnicki}}, \bibinfo{journal}{arXiv
  preprint arXiv:2211.09157}  (\bibinfo{year}{2022}),
  \urlprefix\url{https://arxiv.org/abs/2211.09157}.

\bibitem[{\citenamefont{Zhou and Jiang}(2019)}]{Modern}
\bibinfo{author}{\bibfnamefont{S.}~\bibnamefont{Zhou}} \bibnamefont{and}
  \bibinfo{author}{\bibfnamefont{L.}~\bibnamefont{Jiang}},
  \bibinfo{journal}{Phys. Rev. A} \textbf{\bibinfo{volume}{99}},
  \bibinfo{pages}{013808} (\bibinfo{year}{2019}),
  \urlprefix\url{https://link.aps.org/doi/10.1103/PhysRevA.99.013808}.

\bibitem[{\citenamefont{Tsang}(2019{\natexlab{b}})}]{PhysRevA.99.012305}
\bibinfo{author}{\bibfnamefont{M.}~\bibnamefont{Tsang}},
  \bibinfo{journal}{Phys. Rev. A} \textbf{\bibinfo{volume}{99}},
  \bibinfo{pages}{012305} (\bibinfo{year}{2019}{\natexlab{b}}),
  \urlprefix\url{https://link.aps.org/doi/10.1103/PhysRevA.99.012305}.

\bibitem[{\citenamefont{Tsang}(2021)}]{PhysRevA.104.052411}
\bibinfo{author}{\bibfnamefont{M.}~\bibnamefont{Tsang}},
  \bibinfo{journal}{Phys. Rev. A} \textbf{\bibinfo{volume}{104}},
  \bibinfo{pages}{052411} (\bibinfo{year}{2021}),
  \urlprefix\url{https://link.aps.org/doi/10.1103/PhysRevA.104.052411}.

\bibitem[{\citenamefont{Tan and Tsang}(2023)}]{TsangIII}
\bibinfo{author}{\bibfnamefont{X.-J.} \bibnamefont{Tan}} \bibnamefont{and}
  \bibinfo{author}{\bibfnamefont{M.}~\bibnamefont{Tsang}},
  \bibinfo{journal}{arXiv preprint arXiv:2308.04317}  (\bibinfo{year}{2023}),
  \urlprefix\url{https://arxiv.org/abs/2308.04317}.

\bibitem[{\citenamefont{\ifmmode \check{R}\else
  \v{R}\fi{}eha\ifmmode~\check{c}\else \v{c}\fi{}ek
  et~al.}(2017)\citenamefont{\ifmmode \check{R}\else
  \v{R}\fi{}eha\ifmmode~\check{c}\else \v{c}\fi{}ek, Hradil, Stoklasa, Pa\'ur,
  Grover, Krzic, and S\'anchez-Soto}}]{PhysRevA.96.062107}
\bibinfo{author}{\bibfnamefont{J.}~\bibnamefont{\ifmmode \check{R}\else
  \v{R}\fi{}eha\ifmmode~\check{c}\else \v{c}\fi{}ek}},
  \bibinfo{author}{\bibfnamefont{Z.}~\bibnamefont{Hradil}},
  \bibinfo{author}{\bibfnamefont{B.}~\bibnamefont{Stoklasa}},
  \bibinfo{author}{\bibfnamefont{M.}~\bibnamefont{Pa\'ur}},
  \bibinfo{author}{\bibfnamefont{J.}~\bibnamefont{Grover}},
  \bibinfo{author}{\bibfnamefont{A.}~\bibnamefont{Krzic}}, \bibnamefont{and}
  \bibinfo{author}{\bibfnamefont{L.~L.} \bibnamefont{S\'anchez-Soto}},
  \bibinfo{journal}{Phys. Rev. A} \textbf{\bibinfo{volume}{96}},
  \bibinfo{pages}{062107} (\bibinfo{year}{2017}),
  \urlprefix\url{https://link.aps.org/doi/10.1103/PhysRevA.96.062107}.

\bibitem[{\citenamefont{Huang et~al.}(2023)\citenamefont{Huang, Schwab, and
  Lupo}}]{astro22}
\bibinfo{author}{\bibfnamefont{Z.}~\bibnamefont{Huang}},
  \bibinfo{author}{\bibfnamefont{C.}~\bibnamefont{Schwab}}, \bibnamefont{and}
  \bibinfo{author}{\bibfnamefont{C.}~\bibnamefont{Lupo}},
  \bibinfo{journal}{Phys. Rev. A} \textbf{\bibinfo{volume}{107}},
  \bibinfo{pages}{022409} (\bibinfo{year}{2023}),
  \urlprefix\url{https://link.aps.org/doi/10.1103/PhysRevA.107.022409}.

\end{thebibliography}

\providecommand{\noopsort}[1]{}\providecommand{\singleletter}[1]{#1}%

\end{document}